\documentstyle[aps,prl,multicol,epsf]{revtex}

\begin{document}   
\draft

\title {Scaling Approach to Calculate
Critical Exponents in Anomalous Surface Roughening}

\author{ 
Juan M. L\'opez\cite{mail} 
}

\address{Department of Mathematics,   
Imperial College, London SW7 2BZ, United Kingdom}                       
   
\maketitle    
    
\begin{abstract}   
We study surface growth models exhibiting anomalous
scaling of the local surface fluctuations.
An analytical approach 
to determine the local scaling exponents of 
continuum growth models is proposed.
The method allows to predict when a particular growth model
will have anomalous properties ($\alpha \neq \alpha_{loc}$)
and to calculate the local exponents. Several continuum growth 
equations are examined as examples.
\end{abstract}   

\pacs{64.60.Ht,05.40.+j,61.50.Cj,05.70.Ln} 

\begin{multicols}{2}
\narrowtext     
       
Kinetic roughening of surfaces in nonequilibrium conditions
has been a subject of great interest in the past few years.
This is mainly due to the many important applications 
of the theory of surface growth including molecular-beam
epitaxy (MBE), fluid flow in porous media and fracture cracks
among others \cite{barabasi}.
Growth processes have very often 
been shown to exhibit scaling properties
that allow one to divide the models into universality
classes characterized by the value of the critical 
exponents \cite{barabasi,krug-rev}. 
The {\em global} interface width 
in a system of total lateral size $L$,
which is the root mean square of the 
fluctuations of the surface height, 
scales according to the Family-Vicsek ansatz \cite{fv} as
\begin{equation}
\label{FV-globalwidth}
W(L,t) = t^{\alpha/z} f(L/t^{1/z}),
\end{equation}
where the scaling function 
$f(u)$ behaves as
\begin{equation}
\label{FV-forf}
f(u) \sim
\left\{ \begin{array}{lcl}
     u^{\alpha}     & {\rm if} & u \ll 1\\
     {\rm const.} & {\rm if} & u \gg 1
\end{array}
\right..
\end{equation}
The roughness exponent $\alpha$ and the dynamic
exponent $z$ characterize the universality class
of the model under study. The ratio $\beta = \alpha/z$
is the time exponent. 

However, very recent numerical studies have revealed
a rich variety of interesting phenomena, which are
poorly understood at present. In particular,
it has been shown that many growth models 
\cite{schro,krug,lopez96,dassarma,lopez97a,lopez97b,mario,dasgupta} 
exhibit {\em anomalous} roughening, {\it i.e.}
a different scaling for the global and the local 
surface fluctuations. This leads to the existence
of an independent {\em local} roughness 
exponent $\alpha_{loc}$ that characterizes 
the local interface fluctuations 
on scales $l \ll L$ and differs from $\alpha$. 

A general analytical method to determine
anomalous exponents from the continuum growth 
equations is still lacking. In particular, 
a dynamic renormalization-group (RG) calculation
in the Lai, Das Sarma and Villain (LDV) nonlinear
continuum model has had only
limited success \cite{infrared} because
of the existence of nonperturbative infrared singularities
that are inaccessible to the usual dynamical RG analysis.
Therefore, most efforts have been focused on
determining the anomalous critical 
exponents using simulation of discrete models or direct numerical 
solutions of the Langeving-type equations of growth.

In this Letter, we propose a new analytical approach 
to determine the scaling exponents of the local
surface fluctuations in 
continuum growth models exhibiting anomalous kinetic roughening.
Our method allows to predict when a particular growth model
is expected to have anomalous properties ($\alpha \neq \alpha_{loc}$)
and, in principle, also to determine the local exponents.
We illustrate the method by studying several
growth equations with and without anomalous scaling.

We shall be interested here in continuum growth models
in $d+1$ dimensions which dynamics is expected to be described
by the Langevin-type equation
\begin{equation}
\label{langevin}
{\partial h \over \partial t} = \Phi(\nabla h) + \eta({\bf x},t),
\end{equation}
where $h({\bf x},t)$ is the height of the interface at 
substrate position ${\bf x}$ at time $t$. 
The actual form of the function
$\Phi(\nabla h)$ defines a particular model \cite{phi}.
$\eta({\bf x},t)$ is a noise term     
uncorrelated in space and time, 
$\langle \eta({\bf x},t) \eta({\bf x}',t')\rangle 
= 2D \delta({\bf x}-{\bf x}') \delta(t-t')$. 

The scaling properties of the local surface fluctuations
can be investigated by computing either
the height-height correlation function, 
$G(l,t) = \langle \overline{[h({\bf x}+{\bf l},t) 
- h({\bf x},t)]^2}\rangle$,
where the average is calculated over all ${\bf x}$ (overline) and 
noise (brackets), 
or the {\em local} width, 
$w(l,t) = \langle \langle [h({\bf x},t) - 
\langle h \rangle_l]^2\rangle_l\rangle^{1/2}$, where
$\langle \cdots \rangle_{l}$ denotes an average over ${\bf x}$ 
in windows of size $l$. 
For growth processes in which 
an anomalous roughening takes place these functions 
scale as 
\begin{equation}
\label{anom-scal}
w(l,t) \sim \sqrt{G(l,t)} = t^{\beta} f_A(l/t^{1/z}), 
\end{equation}
with an anomalous scaling 
function \cite{dassarma,lopez97a,lopez97b} given by
\begin{equation}
\label{f_A}
f_A(u) \sim
\left\{ \begin{array}{lcl}
     u^{\alpha_{loc}}     & {\rm if} & u \ll 1 \\
     {\rm const.} & {\rm if} & u \gg 1 
\end{array}
\right., 
\end{equation}
instead of Eq.(\ref{FV-forf}). 
The standard self-affine
Family-Vicsek scaling \cite{fv} is then recovered when 
$\alpha = \alpha_{loc}$. 

It is important to note that
anomalous scaling stems from the fact that the 
mean square local slope has a non-trivial dynamics.
A standard (Family-Vicsek)
self-affine scaling of 
the local interface fluctuations, 
{\it i.e.} $\alpha=\alpha_{loc}$, 
implies that the square local
slope $G(l=a,t) = 
\langle \overline{[h({\bf x}+{\bf a},t)-h({\bf x},t)]^2}\rangle$,
where $a$ is the lattice spacing, 
becomes $G(a,t) \sim {\rm constant}$ very rapidly in time. 
One can also see that this constant must go to 
zero as $a \to 0$.
However, as can be easily seen from Eqs.(\ref{anom-scal}) and 
(\ref{f_A}), in growth 
models exhibiting anomalous scaling, 
the local slopes scale as 
\begin{equation}
\label{local-slopes}
G(a,t) \sim t^{2\kappa},
\end{equation}
where the exponent $\kappa =\beta - \alpha_{loc}/z > 0$.
The existence of such a 
diverging mean local slope 
introduces a new correlation length in the growth direction,
which enters the scaling of the local fluctuations of the height.
Therefore, in any growth model,
the existence (or absence) of anomalous scaling 
of the local height fluctuations 
can be investigated by computing the mean
local slope $G(a,t)$. In the continuous limit, $a \to 0$,
$G(a,t)$ can be written as $G(a,t) \simeq s(t) a^2$ where
\begin{equation}
\label{s}
s(t) = \langle \overline{(\nabla h)^2} \rangle.
\end{equation}
This corresponds to calculating the mean 
square local derivative of the interface height. 
In general, this quantity scales 
as a power law $s(t) \sim t^{2\kappa}$. 
Negative values
of the exponent $\kappa$ will result in a normal 
Family-Vicsek scaling of the local fluctuations 
with the same roughness exponent. 
On the contrary, for $\kappa > 0$ 
the correlation length $s(t)$ diverges 
in time and becomes a relevant length scale in the problem
that changes the local scaling, as discussed above. In this
case, anomalous scaling with a local roughness exponent 
\begin{equation}
\label{alfa-loc}
\alpha_{loc} = \alpha - z \kappa,
\end{equation}
is expected to occur.
As we will see in the following, 
this simple observation allow us to find a general 
method to compute anomalous critical exponents. 
Note also that the exponent $\kappa$ corresponds to the anomalous 
time exponent
$\beta_*$ in Ref.\cite{lopez96,lopez97a,lopez97b}.

Now, let us see how the scaling
behaviour of the mean local slope $s(t)$ 
can be obtained from
the continuum growth equation.
By applying the operator $\nabla$ to both sides
of the growth equation (\ref{langevin}) one gets to
an equation of motion for the local derivative 
$\Upsilon({\bf x},t) = \nabla h$ of the form
\begin{equation}
\label{tangent}
{\partial \Upsilon \over \partial t} =
{\delta \Phi \over \delta \Upsilon} \nabla \Upsilon + \nabla \eta.
\end{equation}
In general, this Langevin equation may contain 
nonlinear terms that break the translational symmetry 
$h \to h + c$ in the growth direction. This implies that the
resulting interface $\Upsilon({\bf x},t)$ may not be  
rough ($\alpha = 0$). 

The global width of the 
interface $\Upsilon({\bf x},t)$ is given by
\begin{equation}
\label{W_U}
W_\Upsilon(t) = \langle \overline{\Upsilon({\bf x},t)^2}\rangle^{1/2}
= \langle \overline{(\nabla h)^2}\rangle^{1/2}
\end{equation}
where $\langle \overline{\nabla h} \rangle = 0$
has been used. 
This leads to the general result that $s(t) = W_\Upsilon^2(t)$ and 
the exponent $\kappa$ of the average local 
interface slope $s(t) \sim t^{2\kappa}$ is given by
the time exponent of the 
global width of the local derivative $\Upsilon({\bf x},t)$,
\begin{equation}
\label{kappa}
W_\Upsilon(t) \sim t^{\kappa}.
\end{equation}
It then becomes clear that 
one could obtain the anomalous 
exponents of the interface $h({\bf x},t)$
by finding the time scaling
behaviour of the fluctuations of $\Upsilon = \nabla h$.

In the following we investigate 
the existence of anomalous scaling in several continuum growth
models. In the examples that we analyse here, a Flory-type 
approach introduced by Hentschel and Family 
\cite{flory} suffices to obtain the exponent $\kappa$ 
from the corresponding Eq.(\ref{tangent}) for every model
in good agreement with existing simulation results.
The Flory approach can be seen as a stability analysis of
the equation of motion (\ref{tangent}) for the corresponding
local derivative $\Upsilon({\bf x},t)$ of the interface.

{\em Kardar-Parisi-Zhang equation.-}
The first system we examine is the noise-driven 
interface growth model given by the 
Kardar-Parisi-Zhang (KPZ) equation
\cite{kpz} 
\begin{equation}
\label{kpz-eq}
{\partial h \over \partial t} = 
\nu \nabla^2 h 
+ \lambda \left(\nabla h \right)^2
+ \eta({\bf x},t).
\end{equation}
This equation was originally introduced 
to describe ballistic deposition growth far from equilibrium.
In $1+1$ dimensions the critical exponents $\alpha=1/2$ and $z=3/2$
can be calculated exactly \cite{kpz}.
On the basis of much numerical work,
it is well established that the KPZ equation does not
exhibit anomalous roughening. Let us see how this result
can be derived 
in dimension $1+1$ by use of our approach.

In this case, the growth
equation for the local derivative, Eq.(\ref{tangent}), reads
\begin{equation}
\label{kpz-tang}
{\partial \Upsilon \over \partial t} =
\nu {\partial^2 \Upsilon \over \partial x^2 }
+ \lambda {\partial \over \partial x}\Upsilon^2 
+ {\partial \eta \over \partial x},
\end{equation}
in $1+1$ dimensions. From this growth equation 
the scaling behaviour of the fluctuations of the
interface $\Upsilon(x,t)$ can be obtained. 
We find that the width scales as 
$W_\Upsilon(t) \sim t^{-1/5}$, where the exponent
can be obtained by a Flory-type approach \cite{flory}
as follows.

In the spirit of Ref.\cite{flory}, we assume that
at long times $t \gg t_l$, and averaged over 
length scales $l$, the typical magnitude of the 
fluctuations in $\Upsilon(x,t)$ scale as $\Upsilon_l$,
and that these fluctuations last for times of the 
order $t_l$. The idea now is to estimate the
magnitude of the individual terms. Basically for
any equation such as Eq.(\ref{kpz-tang}) to
show time scaling each separate term, when coarse-grained
over length scales $l$, must be of the same order of 
magnitude or negligible. Only under these circumstances 
can scaling behaviour arise. The various
terms in Eq.(\ref{kpz-tang}) may be estimated as
$\langle |\partial \Upsilon/\partial t|\rangle_l 
\sim \Upsilon_l/t_l$,
$\langle |\partial^2 \Upsilon/\partial x^2| \rangle_l 
\sim \Upsilon_l/l^2$,
$\langle |\partial \Upsilon^2/\partial x|\rangle_l
\sim \Upsilon_l^2/l$. 
As for the noise, one can 
estimate its fluctuations on lenght scales $l$
and time scales $t_l$ as 
$\langle | \partial \eta/\partial x|\rangle_l
\sim (l^3 t_l)^{-1/2}$
for smooth surfaces, 
while for rough surfaces 
$\langle | \partial \eta/\partial x|\rangle_l
\sim (\Upsilon_l l^2 t_l)^{-1/2}$ (see Ref.\cite{flory}
for details). Note that the correct 
scaling of the noise term depends
on the particular equation under study. 

At sufficiently large length
scales the nonlinear term in Eq.(\ref{kpz-tang}) will
dominate the diffusion term, and so, equating the
$\langle |\partial \Upsilon/\partial t|\rangle_l$ term
with the nonlinear term we obtain that a typical 
fluctuation scales as $\Upsilon_l \sim l/t_l$.
To proceed further we now equate our estimate for the
noise fluctuation $(\Upsilon_l l^2 t_l)^{-1/2}$
to the inertial term, which gives
$\Upsilon_l \sim (t_l/l^2)^{1/3}$. So, we can estimate
that a fluctuation of $\Upsilon$ scales in time as
$\Upsilon_l \sim t_l^{-1/5}$. 

So, in the case of the KPZ equation in $1+1$ dimensions 
a negative exponent $\kappa = -1/5$ is found indicating
a standard scaling as expected. A similar 
Flory-type computation also shows that, in fact, 
the KPZ model exhibits a
self-affine scaling in $d+1$ dimensions 
and $\kappa = - 1/(4+d)$.

A particular case of the KPZ equation is the 
linear interface 
growth model ($\lambda = 0$), first studied by
Edwards and Wilkinson \cite{ew}. For this model 
Eq.(\ref{kpz-tang}) can be solved exactly and we
obtain the exponent $\kappa = -1/4$, as corresponds
to a standard Family-Vicsek scaling.

{\em Surface growth with conservation law.-}
The KPZ equation does not conserve the total volume
of the interface. The conserved version of KPZ was
studied by Sun, Guo and Grant \cite{sgg} 
and is given by
\begin{equation}
\label{sgg-eq} 
{\partial h \over \partial t} = 
- K \nabla^4 h
+ \lambda \nabla^2  
\left(\nabla h \right)^2 
+ \eta_c({\bf x},t),
\end{equation}
where 
\begin{equation}
\langle \eta_c({\bf x},t) \eta_c({\bf x}',t')\rangle = 
- 2D \nabla^2 \delta({\bf x} - {\bf x}') \delta(t-t').
\end{equation}
Here the exponents are exact in any dimension \cite{sgg}, 
in particular $\alpha=1/3$ and $z=11/3$ for $d=1$. 
We have investigated the possibility of anomalous scaling in this
model in $1+1$ dimensions. From the corresponding 
equation for the local derivative
Eq.(\ref{tangent}) and the
noise fluctuations 
$\sim (l^3 t_l)^{-1/2}$,  
we obtain the scaling behaviour
$W_\Upsilon(t) \sim t^{-2/11}$ for the 
fluctuations of $\Upsilon(x,t)$. 
This result shows that
the SGG equation has also a normal scaling of the local
fluctuations of the height.
  
{\em Linear MBE model.-}
As a simple example of anomalous roughening, we now 
consider the linear model for 
MBE growth \cite{groove,super-rough} given by
\begin{equation}
\label{linear-MBE}
{\partial h \over \partial t} = 
- K \nabla^4 h 
+ \eta({\bf x},t),
\end{equation}
Despite its simplicity, 
this equation has played an important
role in the theory of MBE. The critical 
exponents are easily calculated in any dimension,
and in particular one has $\alpha=3/2$ and $z=4$
in dimension $d=1$. The model exhibits super-rough
interfaces ($\alpha > 1$). This leads to anomalous
(super-rough) scaling \cite{lopez97a,lopez97b}. 
In fact, numerical 
simulations \cite{groove,super-rough,lopez97a,lopez97b}
of Eq.(\ref{linear-MBE}) showed that the local scaling is given
by Eqs.(\ref{anom-scal}) and (\ref{f_A}) with a
local roughness exponent $\alpha_{loc} \simeq 1$.

In this case the growth equation for the local derivative
\begin{equation}
\label{mbe-tang}
{\partial \Upsilon \over \partial t} = 
- K {\partial^4 \Upsilon \over \partial x^4 }
+ {\partial \over \partial x} \eta(x,t),
\end{equation}
is linear and can be easily solved. The Flory
approach now gives the exact exponent $\kappa$.
We can estimate the curvature diffusion term as 
$\langle |\partial^4 \Upsilon/\partial x^4| \rangle_l 
\sim \Upsilon_l/l^4$, being the estimate for the
noise term 
$\langle | \partial \eta/\partial x|\rangle_l
\sim (l^3 t_l)^{-1/2}$ in this case.
Equating fluctuations
of each of the two terms on the right-hand side of
Eq.(\ref{mbe-tang}) with the inertial term, 
we obtain the scaling behaviour $W_\Upsilon(t) \sim t^{1/8}$ 
for the width of the local interface derivative. 
The positive value
of the exponent $\kappa$ means that the local slope $s(t)$
becomes a relevant correlation length in the problem and
anomalous roughening is to be expected. 
The scaling relation (\ref{alfa-loc})
gives us an exact determination of the 
local roughness exponent $\alpha_{loc} = 1$ for this model in
dimension $d=1$.
This is good in agreement with existing numerical 
results \cite{groove,super-rough,lopez97a}.

{\em Random diffusion model.-} 
So far we have been considering growth models in which 
the only source of randomness is in the influx of particles
on the surface. Let us now consider the growth model
\begin{equation}
\label{rdmodel}
{\partial h \over \partial t} = 
{\partial \over \partial x} [D(x)
{\partial h \over \partial x}] + \eta(x,t),
\end{equation}
where $D(x) > 0$ is a quenched columnar disorder with
no correlations. This random diffusion 
coefficient is distributed according 
to $P(D) \sim D^{-\mu}$ for $D < 1$
and $P(D) = 0$ for $D > 1$.  
This model describes
a fluid interface advancing through a 
stratified porous medium
with long range correlations in the growth direction. 

The random diffusion model was originally 
introduced \cite{lopez96}
to demonstrate that anomalous roughening is
not due to super-roughening effects, but it can also
take place in models in which $\alpha < 1$.
The critical exponents depend on the disorder 
strength $\mu$ and can be calculated exactly \cite{lopez96}
$\alpha = 1/[2(1-\mu)]$ and $z = (2-\mu)/(1-\mu)$ for 
$0 < \mu < 1$. This model exhibits anomalous roughening
with a local roughness exponent $\alpha_{loc} = 1/2$ 
that is independent of the disorder \cite{lopez96}.

The existence of anomalous roughening in this model 
can be rederived by use of our approach as follows.
The growth equation for the local derivative 
\begin{equation}
{\partial \Upsilon \over \partial t} = 
{\partial^2 \over \partial x^2}[D(x) \Upsilon] 
+ {\partial \over \partial x} \eta(x,t),
\end{equation}
can be solved exactly to obtain the time scaling
behaviour of $W_\Upsilon(t) \sim t^\kappa$. We find
that $\kappa = \mu/[2(2-\mu)]$ for $0 < \mu < 1$, which
gives us a local roughness exponent $\alpha_{loc} = 
\alpha -z \kappa = 1/2$ in agreement with earlier
results \cite{lopez96}.

{\em Lai-Das Sarma-Villain equation.-}
The last example we study is the LDV \cite{ldv,kim}
equation for MBE growth 
\begin{equation}
\label{ldv-eq}
{\partial h \over \partial t} = 
- K \nabla^4 h
+ \lambda \nabla^2 
\left(\nabla h \right)^2 
+ \eta({\bf x},t).
\end{equation}
This equation is expected to describe 
the behaviour of the long-wavelength fluctuations
of the interface height in several discrete models
of MBE growth \cite{krug,dassarma,kim,huang,dasgupta}.
Note that this equation differs from the SGG equation
discussed above in the non-conserved character of
the noise in this case.

According to a dynamical RG analysis of Eq.(\ref{ldv-eq})
the global exponents $\alpha = (4-d)/3$ and $z = (8+d)/3$
are expected to be exact to all loops.
However, numerical simulations of the LDV equation
and several discrete growth models in the same universality
class have shown that the model exhibits anomalous scaling
with a local roughness exponent 
$\alpha_{loc} \simeq 0.7$ \cite{krug,dassarma,dasgupta} in $1+1$ dimensions. 
A perturbative dynamic RG approach 
to Eq.(\ref{ldv-eq}) showed \cite{infrared}
the existence of a strong-coupling infrared 
singularity that prevented from
obtaining the local anomalous behaviour by perturbative
methods.

Again, the existence of anomalous 
scaling in this model in $1+1$ dimensions
can be investigated by use of the growth equation for the 
local derivative of the interface.
In this case Eq.(\ref{tangent}) reads
\begin{equation}
\label{ldv-tang}
{\partial \Upsilon \over \partial t} = 
- K {\partial^4 \Upsilon \over \partial x^4 }
+ \lambda {\partial^3 \over \partial x^3 } 
\left(\Upsilon^2\right)
+ {\partial \over \partial x} \eta(x,t).
\end{equation}
A Flory approach can also be applied to this
case to determine the scaling of $W_\Upsilon \sim t^\kappa$.
As we did for the KPZ equation, 
we can estimate the terms in Eq.(\ref{ldv-tang}) as
$\langle |\partial \Upsilon/\partial t|\rangle_l 
\sim \Upsilon_l/t_l$,
$\langle |\partial^4 \Upsilon/\partial x^4| \rangle_l 
\sim \Upsilon_l/l^4$,
$\langle |\partial^3 \Upsilon^2/\partial x^3|\rangle_l
\sim \Upsilon_l^2/l^3$, and the noise being
$\langle | \partial \eta/\partial x|\rangle_l 
\sim (\Upsilon_l l^2 t_l)^{-1/2}$ as in the KPZ case.
Assuming that the nonlinear term dominates the curvature
diffusion term at large scales, we obtain 
$W_\Upsilon \sim t^{1/11}$.

According to this the exponent $\kappa = 1/11$ is positive and,
as a consequence, the LDV equation in $1+1$ dimensions
is expected to display anomalous roughening. The
local roughness exponent $\alpha_{loc} = 8/11 \simeq 0.73$
is given by the scaling relation (\ref{alfa-loc}) with
$\alpha=1$ and $z=3$ for $d=1$. 
Our estimate for $\alpha_{loc}$ is in excellent
agreement with the numerical result 
$\alpha_{loc} \simeq 0.73 \pm 0.04$ \cite{krug}.

{\em Conclusion.-}
In summary, we have introduced a new method to calculate
the local roughness exponent in growth models 
exhibiting anomalous kinetic roughening.
We have shown that a divergent
dynamics of the mean square local derivative, 
$s(t) \sim t^{2\kappa}$ with $\kappa > 0$,
is the cause for an anomalous roughening of the local surface 
fluctuations. The exponent $\kappa$ can be obtained by
studying the time scaling behaviour of the fluctuations of the
local derivative. The local roughness exponent can then be obtained
by use of a scaling relation, Eq.(\ref{alfa-loc}).
We have examined the existence of anomalous
roughening in several models. For the linear models our results
are exact. Nonlinear growth equations were studied by a 
scaling approach, which gave results in good agreement with
existing simulations. 

Finally, our findings indicate that a self-affine ($\alpha=\alpha_{loc}$)
Family-Vicsek scaling of the interface fluctuations is the result
of a fine balance among the relevant terms in the growth equation.
This balance can be altered by including conservation laws or 
higher order nonlinear terms resulting in a divergent behaviour
of the local derivatives $s(t)$ and anomalous scaling.

The author 
would like to thank Miguel A. Rodr{\'\i}guez for many
interesting discussions and a careful reading of the
manuscript. This work is supported by the
European Commission.

\end{multicols}

\end{document}